\documentclass[11pt]{gh2013}

\usepackage{mathptmx}       
\usepackage{helvet}         
\usepackage{courier}        
\usepackage{makeidx}         
\usepackage{graphicx}        
\usepackage{multicol}        
\newcommand{\oiii}{[O\thinspace{\sc iii}]}
\newcommand{\heii}{He\thinspace{\sc ii}}
\newcommand{\hi}{H\thinspace{\sc i}}
\newcommand{\oii}{[O\thinspace{\sc ii}]}
\newcommand{\oi}{[O\thinspace{\sc i}]}
\newcommand\fesc{$f_{\rm esc}$}
\newcommand{\Msun}{M$_{\odot}$}
\newcommand{\Zsun}{Z$_{\odot}$}
\newcommand{\cii}{C\thinspace{\sc ii}}
\newcommand{\sitwo}{Si\thinspace{\sc ii}}

\begin{document}

\title*{The Origin and Optical Depth of Ionizing Photons in the Green Pea Galaxies}
\titlerunning{Ionization in the Green Pea Galaxies}
\author{A. E. Jaskot$^{1}$ and M. S. Oey$^{1}$}
\institute{$^{1}$Department of Astronomy, University of Michigan, Ann Arbor, MI 48109-1042, USA}
%
%
\maketitle


\vskip -3.5 cm  
\abstract{
Our understanding of radiative feedback and star formation in galaxies at high redshift is hindered by the rarity of similar systems at low redshift. However, the recently identified Green Pea (GP) galaxies are similar to high-redshift galaxies in their morphologies and star formation rates and are vital tools for probing the generation and transmission of ionizing photons. The GPs contain massive star clusters that emit copious amounts of high-energy radiation, as indicated by intense \oiii~$\lambda$5007 emission and \heii~$\lambda$4686 emission. We focus on six GP galaxies with high ratios of \oiii~$\lambda\lambda$5007,4959/\oii~$\lambda$3727 $\sim$10 or more. Such high ratios indicate gas with a high ionization parameter or a low optical depth. The GP line ratios and ages point to chemically homogeneous massive stars, Wolf-Rayet stars, or shock ionization as the most likely sources of the \heii~emission. Models including shock ionization suggest that the GPs may have low optical depths, consistent with a scenario in which ionizing photons escape along passageways created by recent supernovae. The GPs and similar galaxies can shed new light on cosmic reionization by revealing how ionizing photons propagate from massive star clusters to the intergalactic medium.
}

\section{The Optical Depths of the Extreme Green Peas}
The ionizing photons produced by massive stellar clusters in the early Universe are likely responsible for cosmic reionization. Determining Lyman continuum (LyC) escape fractions (\fesc) from galaxies and the factors that promote LyC escape is therefore critically important. However, the low-luminosity galaxies that dominate the ionizing flux at the redshift of reionization ($z>6$) are unobservable by existing telescopes \cite{Bouwens2011} and their \fesc~values are highly uncertain. Direct detections of LyC at intermediate redshift ($z\sim$3) imply \fesc$=10-20$\% (e.g., \cite{Steidel2001, Iwata2009}), but an unknown fraction of these detections may arise from contamination by low-redshift galaxies along the line of sight \cite{Vanzella2012}. Finally, at $z\sim$0, where we can study galaxy properties in detail, only two galaxies have confirmed LyC detections \cite{Leitet2013}. To understand LyC escape, we need to identify more optically thin galaxies at low redshift.

The Green Pea (GP) galaxies may constitute some of these elusive low-redshift galaxies with high \fesc. Identified in the Sloan Digital Sky Survey (SDSS) by their compact sizes and intense \oiii~$\lambda$5007 emission \cite{Cardamone2009}, the $z=0.1-0.3$ GPs share many properties with high-redshift galaxies. The GPs' enormous \oiii~equivalent widths ($\sim$1000 \AA), high specific star formation rates ($10^{-7}$ yr$^{-1}$), low stellar masses (10$^8$-10$^9$ \Msun), low metallicities ($\sim$0.2 \Zsun), and clumpy morphologies \cite{Cardamone2009, Izotov2011} are similar to the characteristics of galaxies at $z=2-6$ (e.g., \cite{Maseda2013, Nakajima2013, Smit2013, Stark2009}). Most importantly, many GPs have extremely high ratios of \oiii~$\lambda$5007/\oii~$\lambda$3727, implying either a high ionization parameter or a low optical depth (e.g., \cite{Giammanco2005}; see also the presentation by M. S. Oey). These high \oiii/\oii~ratios are comparable to the observed ratios in LyC-leaking galaxies at both low and high redshift \cite{NO2013}.

To establish whether the GPs are optically thin, we investigated whether the galaxies' ionizing sources could reproduce the high observed \oiii/\oii~ratios without the need for a low optical depth \cite{Jaskot2013}. In this analysis, we focused on six ``extreme" GPs with the highest ratios of \oiii/\oii. Five of the six extreme GPs exhibit \heii~$\lambda$4686 emission in their SDSS spectra, which sets strong constraints on the possible ionizing sources. Doubly-ionized He requires 54 eV; the spectral energy distributions of typical massive main-sequence stars are not hard enough to generate significant nebular \heii~emission. Alternative sources of \heii~include Wolf-Rayet (WR) stars, low-metallicity O stars, active galactic nuclei (AGN), high-mass X-ray binaries (HMXBs), and shocks (see \cite{Shirazi2012} and talk by E. Telles for a summary). 

We evaluated these possible ionizing sources based on the GPs' nebular line ratios, \heii~luminosities, inferred burst ages of 3-5 Myr, and estimated starburst masses \cite{Jaskot2013}. The width of the GPs' \heii~emission lines eliminates a WR stellar wind origin, and the observed line ratios and luminosities are inconsistent with the AGN and HMXB scenarios. Photoionization by WR stars is possible; however, the SDSS spectra show no WR features, such as the characteristic ``blue bump" emission at 4640-4686 \AA. Nevertheless, WR features may be weaker at low metallicity and the SDSS spectra detection limits do not rule out the WR photoionization hypothesis. Furthermore, Starburst99 models \cite{Leitherer1999} indicate that a 4-5 Myr stellar population could produce the observed \heii. A second possibility is photoionization by hot, rapidly rotating, chemically homogeneous O stars, which may be more common at low metallicities (e.g., \cite{Brott2011, Maeder1987}). However, these stars may have strong WR features \cite{Eldridge2012}, and as observations have not yet identified stars of this type, we cannot evaluate their viability as ionizing sources in the GPs. Finally, shocks may cause the observed \heii, and the GPs do show evidence for shock emission. Many GPs appear to be galaxy mergers \cite{Cardamone2009}, and recent studies demonstrate that shocks may produce a significant fraction of the observed nebular emission in mergers \cite{Krabbe2014, Rich2014}. The broad, high-velocity wings in the GPs' Balmer emission lines \cite{Amorin2012} and the detection of \oi~$\lambda$6300 support the shock scenario, and the first supernovae at 3.5 Myr after the initial burst could account for the observed \heii~\cite{Jaskot2013}.

\begin{figure*}
  \centering
  \includegraphics[width=7.5 cm]{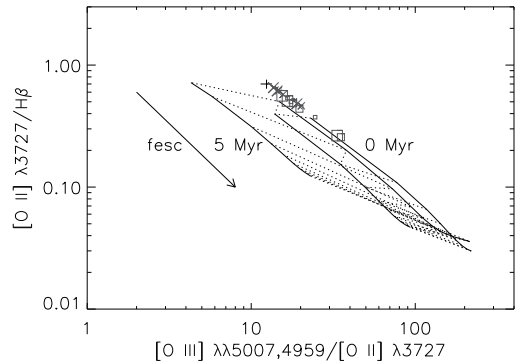}
  \caption{An example of the \oiii/\oii~and \oii/H$\beta$~ratios predicted from CLOUDY photoionization models. The solid lines show the effect of changing the burst age, using Starburst99 instantaneous bursts of 0, 3, 4, and 5 Myr in age. The dashed lines show the line ratios for various optical depths. The uppermost line indicates \fesc=0, and \fesc~increases in the direction shown by the arrow. The black cross shows the observed ratios for one of the extreme GPs, and the other symbols show how the inferred ratios change after subtracting the shock contribution using different Mappings III shock models. Subtracting shock emission lowers the inferred optical depth. (Adapted from \cite{Jaskot2013}.) }
  \label{fig:shocks}
\end{figure*}

We now consider what the potential ionizing sources imply for the GPs' optical depths. Using CLOUDY photoionization models \cite{Ferland1998} with Starburst99 instantaneous bursts \cite{Leitherer1999} as the ionizing sources, we find that an optically thick nebula ionized by a $\sim$4 Myr burst containing WR stars could match the high observed \oiii/\oii~ratios \cite{Jaskot2013}. However, if shocks, not WR stars, cause the \heii~ionization, then we need to account for the shock contribution to other emission lines, such as \oii. Subtracting the shock contribution from the observed nebular emission using Mappings III shock models \cite{Allen2008} results in even higher \oiii/\oii~ratios in the photoionized gas and implies a low optical depth \cite{Jaskot2013} (Fig.~\ref{fig:shocks}).

\begin{figure*}
  \centering
  \includegraphics[width=6.0 cm]{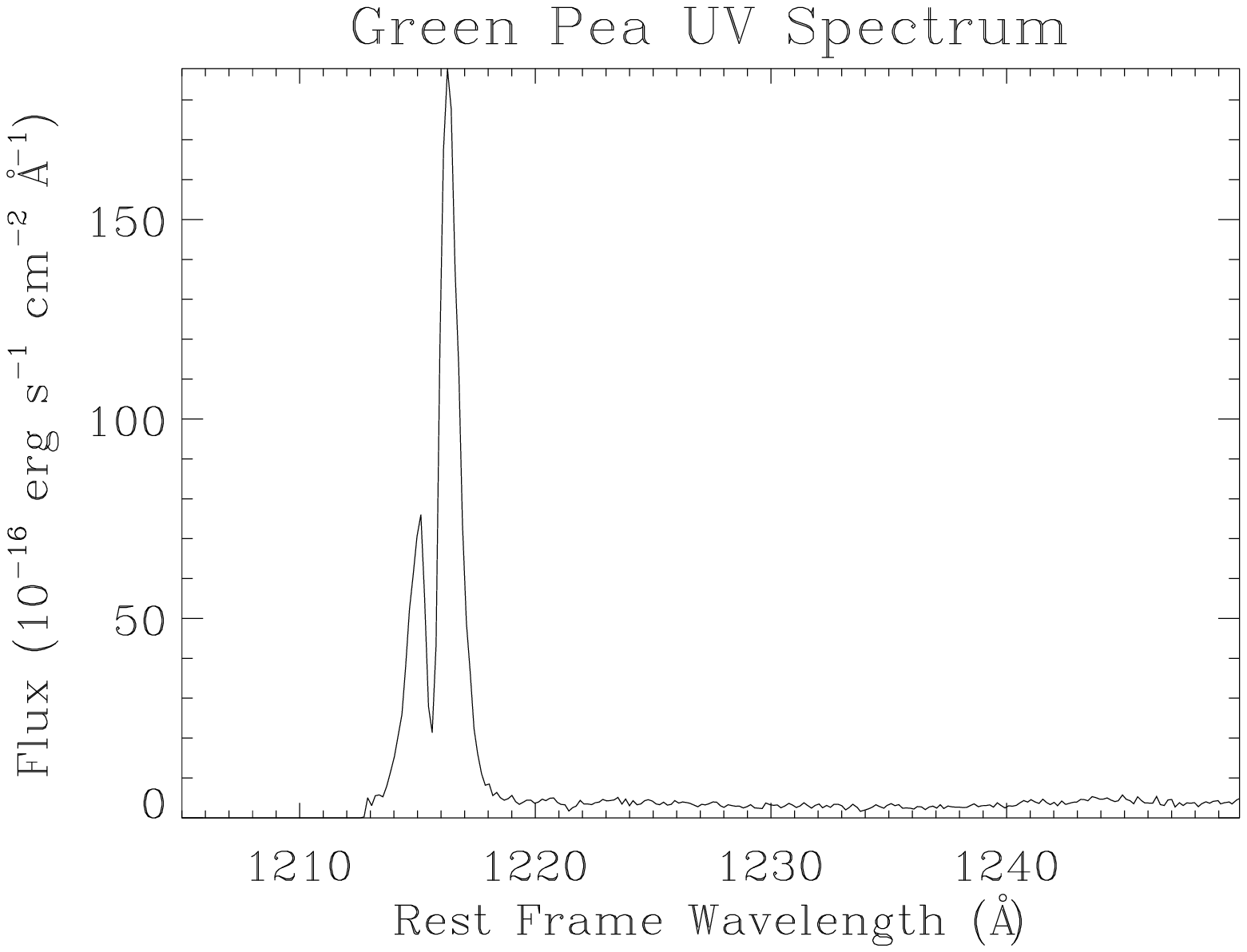}
  \includegraphics[width=6.0 cm]{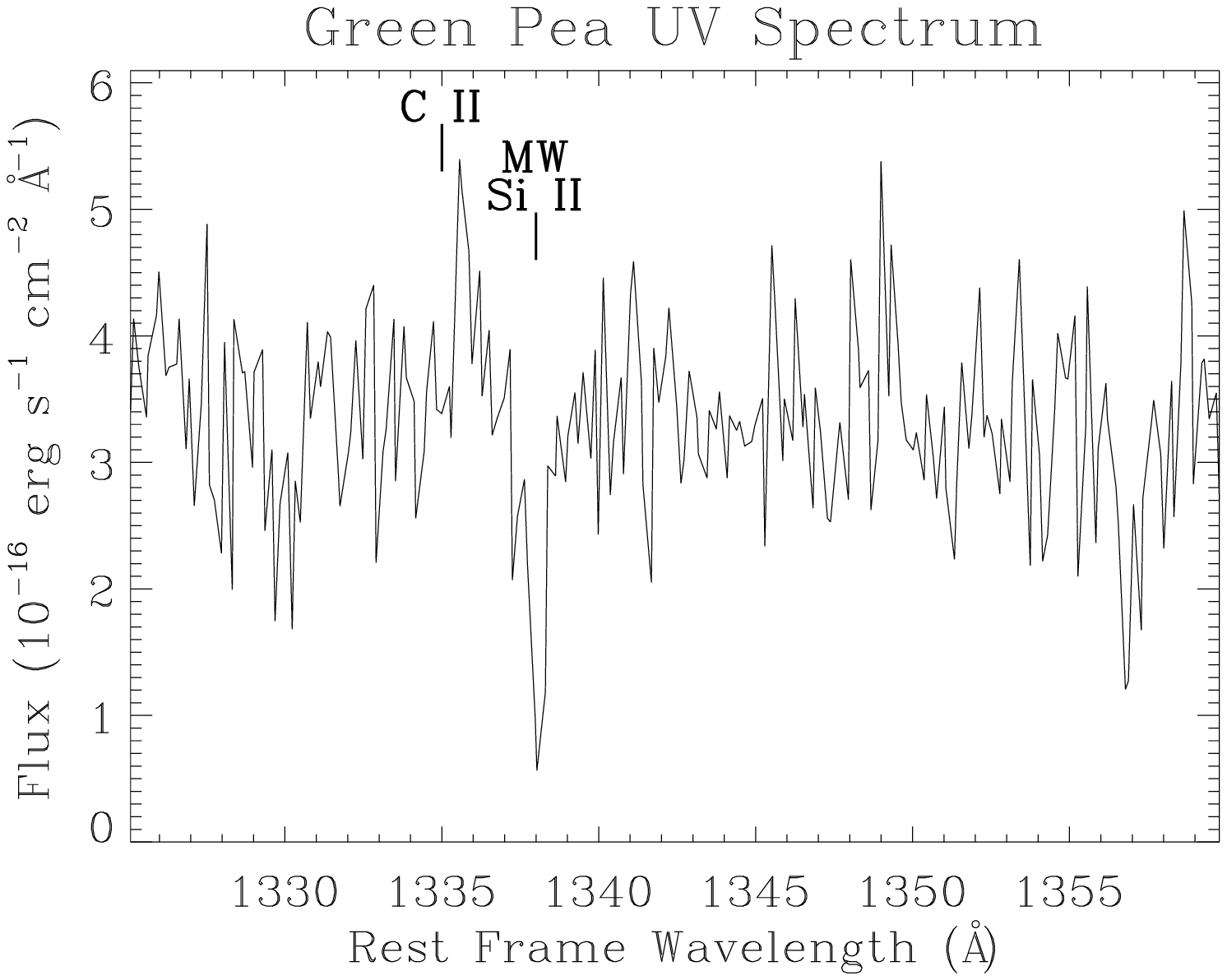}
  \caption{COS spectra of GP J081552.00+215623.6 showing strong Ly$\alpha$~emission (left) and no \cii~$\lambda$1335 absorption (right). The position of Milky Way \sitwo~absorption is also marked on the spectrum. The lack of \cii~absorption may indicate a low optical depth in the neutral ISM.}
  \label{fig:cos}
\end{figure*}

We are currently analyzing observations from IMACS and MagE on the Magellan telescopes and COS and ACS on {\it HST} to distinguish between the WR and shock ionization scenarios and confirm the GPs' optical depths. The absence of WR features in the deeper IMACS spectra tentatively supports the shock scenario, although the detection limits do not yet definitively rule out the WR photoionization hypothesis. The MagE spectra will settle the question of whether WR stars are present and reveal weak shock lines, if they exist. Our latest {\it HST} observations (GO 13923; P.I. Jaskot) demonstrate that the extreme GPs are Ly$\alpha$~emitters (LAEs), with the strongest Ly$\alpha$~emission present in objects that lack absorption lines from the neutral interstellar medium (ISM), such as \cii~$\lambda$1335 (Fig.~\ref{fig:cos}). The absence of strong \cii~absorption implies that these GPs may be optically thin along our line of sight. The likelihood of shock ionization in the GPs combined with the Ly$\alpha$~emission suggests a possible explanation for low optical depths in the GPs. Supernova feedback 3-5 Myr after a burst may create holes in the ISM, allowing LyC photons from the remaining massive stars to escape \cite{Clarke2002, Zastrow2013}. Supernova-driven outflows should likewise enhance Ly$\alpha$~escape \cite{MasHesse2003}, consistent with our {\it HST} observations and as B. James discussed for Haro 11, Ly$\alpha$~emission may be associated with low \hi~column densities. Burst age and star formation history may therefore play a key role in shaping the neutral ISM and causing LyC escape.


\begin{acknowledgement}
I thank the conference organizers for the opportunity to present this research and acknowledge support from an NSF Graduate Research Fellowship and NASA HST GO-13293.01-A. 

\end{acknowledgement}

\end{document}